\documentclass[aps,prl,twocolumn,superscriptaddress]{revtex4-1}
\usepackage{graphicx}      % alternative graphics specifications
\usepackage{bm}            % special 'bold-math' package
\usepackage{siunitx}
\usepackage{multirow}
\usepackage{makecell}

\begin{document}
\title{Laser Pulse Compression Using Magnetized Plasmas}

\author{Yuan Shi}
\email{yshi@pppl.gov}
\affiliation{Department of Astrophysical Sciences, Princeton University, Princeton, NJ 08544 USA}
\affiliation{Princeton Plasma Physics Laboratory, Princeton University, Princeton, NJ 08543 USA}

\author{Hong Qin}
\affiliation{Department of Astrophysical Sciences, Princeton University, Princeton, NJ 08544 USA}
\affiliation{Princeton Plasma Physics Laboratory, Princeton University, Princeton, NJ 08543 USA}
\affiliation{School of Nuclear Science and Technology, University of Science and Technology of China, Hefei, Anhui 230026, China}

\author{Nathaniel J. Fisch}
\affiliation{Department of Astrophysical Sciences, Princeton University, Princeton, NJ 08544 USA}
\affiliation{Princeton Plasma Physics Laboratory, Princeton University, Princeton, NJ 08543 USA}

\date           {\today}

\begin{abstract}

Proposals to reach the next generation of laser intensities through Raman or Brillouin backscattering have centered on optical frequencies.  
Higher frequencies are beyond the range of such methods mainly due to the wave damping that accompanies the higher density plasmas necessary for compressing higher frequency lasers. 
However, we find that an external magnetic field transverse to the direction of laser propagation can reduce the required plasma density.
Using parametric interactions in magnetized plasmas to mediate pulse compression both reduces the wave damping and alleviates instabilities,  thereby enabling higher frequency or lower intensity pumps to produce pulses at higher intensity and longer duration.  
In addition to these theoretical advantages, our new method,  in which strong uniform magnetic fields lessen the need for high-density uniform plasmas,  also lessens key engineering challenges, or at least exchanges them for different challenges. 
\end{abstract}

\maketitle
\setlength{\parskip}{0pt}

%%%%%%%%%%%%%%%%%%%%%%%%%%%%%%%%%%%%%%%%%%%%%%%%%%%%%%%%%%%%%%%%%%%%%%%%%%%%%
Extremely high intensity lasers could have manifold applications, such as inertial confinement fusion \cite{Kauffman94} and single molecule imaging \cite{Solem82,*Neutze00,*Hau-Riege07}. 
To achieve extreme intensities, parametric compressions have been proposed using plasmas,  with waves such as the Langmuir wave and the ion acoustic wave mediating the compression \cite{Milroy79,*Capjack82,*Malkin99,Andreev06,*Weber13,*Frank2014,*Lehmann16,Riconda13,*Jia16}. 
At optical frequencies, a window exists in the plasma density-temperature space wherein neither the plasma waves nor the lasers are heavily damped. 
However,  for higher frequency lasers, higher density plasmas are required to mediate the interaction, and at higher density these waves tend to be heavily damped. 
Here we propose,  by utilizing waves in magnetized plasmas,  to extend the frequency and intensity range of laser pulse compression. 
In magnetized plasmas, waves that can be utilized are the electrostatic waves, including hybrid waves and Bernstein waves. 
These waves provide resonances in which contributions from plasma density are partially replaced by more controllable contributions from the external magnetic field. 
The reduced dependence on plasma density alleviates wave damping as well as deleterious instabilities \cite{Clark03,*Malkin07,*Malkin14},  expanding the operation window of pulse compression  to produce output pulses at both higher intensity and longer duration.

\begin{figure}[b]
	\includegraphics[angle=0,width=6cm]{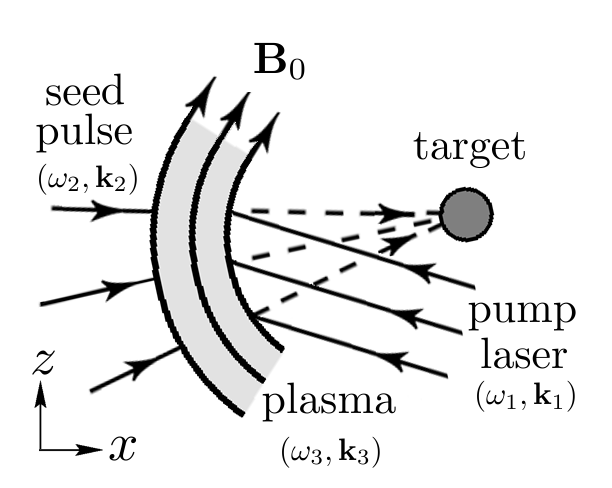}
	\caption{Amplifying and focusing a seed pulse by stimulated backscattering of a pump laser in magnetized plasma.}
	\label{fig:schematics}
\end{figure}

In this letter, we show the advantage of applying a transverse magnetic field by examining pulse compression mediated by the upper-hybrid (\textit{UH}) wave. 
The transverse geometry differs from recent considerations of axial magnetic fields, which affect other aspects of propagation and amplification \cite{Vij16,*Shoucri16,*Luan16}. 
Consider the case where the lasers propagate exactly perpendicular to the external magnetic field, which lends itself naturally to the main application where the amplified pulse is focused onto a distant target (Fig.~\ref{fig:schematics}). 
For propagation perpendicular to the magnetic field, the linear wave eigenmodes are well known \cite{Stix92}. 
One electromagnetic eigenmode is the \textit{O} wave, with electric field  parallel to the external magnetic field, obeying the  dispersion relation $n_\perp^2=1-\omega_p^2/\omega^2$. 
Here $n_\perp=ck_\perp/\omega$ is the refractive index and $\omega_p$ is the plasma frequency. 
The other electromagnetic eigenmode is the \textit{X} wave, which hybridizes with the electrostatic eigenmode, the \textit{UH} wave, obeying the dispersion relation $n_\perp^2=2RL/(R+L)$. 
Here $R,L=1-\omega_p^2/[\omega(\omega\pm\Omega)]$, where $\Omega=eB_0/m_e$ is the electron gyrofrequency. 
The electric field of both the \textit{X} wave and the \textit{UH} wave are in the plane perpendicular to the magnetic field. 
While an \textit{X} wave and an \textit{O} wave couple only weakly, two \textit{X} waves or two \textit{O} waves couple strongly through the \textit{UH} wave.

What we show here is that compression of these electromagnetic waves mediated by the \textit{UH} wave is described by the same equations that describe mediation by the Langmuir wave, except for the coupling coefficient. 
However, because less density is sufficient to accomplish the resonance, key deleterious effects become less competitive.
The result is an expansion of the parameter regime of operation, allowing extension even to the x-ray regime whenever fields of several hundred megagauss to a gigagauss become available.
While such fields are challenging, at the cusp of present feasibility, our new method at least provides a theoretical opportunity to compress lasers that otherwise could not be compressed.  
In addition, at lower frequencies,  more readily available magnetic fields may also confer advantages.

To see this, consider the resonant coupling between the pump laser (with frequency $\omega_1$) and the seed pulse (with frequency $\omega_2$) through the mediating \textit{UH} wave (with frequency $\omega_3=\sqrt{\omega_p^2+\Omega^2}$).
This interaction can be described by the three-wave equations. Using the three-wave resonance conditions, it can be shown that in the limit $\omega_0:=\omega_1\simeq\omega_2\gg\omega_3$, both electromagnetic eigenmodes are transverse with little dispersion, and the \textit{UH} wave is almost longitudinal with approximately zero group velocity. Consequently, the three-wave interaction in the magnetized case has one-to-one correspondence with that in the cold unmagnetized case:
\begin{eqnarray}\label{eq:3-wave}
\nonumber
(\partial_t+c\partial_x)a_1&=&\phantom{-}\frac{\omega_p}{2}a_2 a_3,\\
(\partial_t-c\partial_x)a_2&=&-\frac{\omega_p}{2}a_1 a_3^*,\\
\nonumber
\partial_t a_3&=&-\frac{\omega_0\omega_p}{2\omega_{3}}a_1 a_2^*.
\end{eqnarray}
The electric field amplitude of the pump and the seed are normalized by $a_{1,2}=eE_{1,2}/m_ec\omega_{1,2}$, and the amplitude of the \textit{UH} wave is normalized by $a_3=eE_3/m_ec\omega_p$. The linear growth rate, i.e., for negligible pump depletion, is
\begin{equation}
\Gamma_{R}=\frac{\sqrt{\omega_3\omega_0}}{2}|a_{1}|\gamma_B^{-1},
\end{equation} 
where $\gamma_B:=\omega_3/\omega_p>1$ measures the extent to which plasma density is replaced by magnetic field in the \textit{UH} resonance. 
Except for the above modification, the solution in the nonlinear stage is the same as the unmagnetized cases. 
However, the physical processes that limit pulse compression are different. What is of critical importance is that these physical processes lead to different constraints on the possible amplification regimes.

The first limiting effect is wavebreaking, which limits the maximum pump intensity that can be used for amplifying the pulse. In magnetized plasmas, the wavebreaking intensity is modified by the Lorentz force, and the \textit{UH} wave breaks when the electron quiver velocity in the $\mathbf{k}_3$-direction $v_{q}\simeq eE_3\omega_3/m_e\omega_p^2$ exceeds the wave phase velocity $v_p=\omega_3/k_3\simeq c\omega_3/2\omega_0$  \cite{Karmakar16}. The condition $v_{q}\lesssim v_p$, which guarantees that the \textit{UH} wave remains unbroken, can be rewritten in terms of a constraint on the pump intensity $I_1=8I_c|a_1|^2$ using the Manley-Rowe relation, which constrains the maximum amplitude of the \textit{UH} wave to be $|a_3|\le\sqrt{\omega_0/\omega_3}|a_1|$. The resultant sufficient condition that the \textit{UH} wave remains unbroken is
\begin{equation}\label{eq:WaveBreaking}
I_1\lesssim I_c\Big(\frac{\omega_3}{\omega_0}\Big)^3\gamma_B^{-2},
\end{equation}
where  $I_c=n_cm_ec^3/16$, and $n_c=\epsilon_0m_e\omega_0^2/e^2$ is the critical density. 
When more plasma density is replaced by magnetic field in $\omega_3$, less energy can be contained in the \textit{UH} wave, giving rise to the $\gamma_B^{-2}$ reduction. 
For given laser parameters, wavebreaking constrains the minimum plasma density, as well as the maximum magnetic field.

The second limiting effect is the modulational instability, with growth rate \cite{Malkin99,*Malkin14}
\begin{equation}
	\Gamma_M=\frac{\omega_3^2}{8\omega_0}|a|^2\gamma_B^{-2}. 
\end{equation}
The maximum time that the pulse can be amplified by the pump is limited to a few inverse  growth rates.   
Since $\Gamma_M\ll\Gamma_R$ even at the wavebreaking intensity, this instability does not prevent the amplification from reaching the nonlinear stage, which can continue until
\begin{equation}
	t_M\approx(12\delta\Lambda_0^2)^{1/3}\frac{\gamma_B^{4/3}}{\omega_3a_{10}^{4/3}},
\end{equation}
where $\delta\!=\!\int\!\Gamma_Mdt\!\sim\!1$ is the accumulated phase shift, $\Lambda_0$ is the number of linear exponentiations before the nonlinear stage is reached, and $a_{10}$ is the initial pump amplitude. The largest pulse compression is attained at the maximum compression time $t_M$, which gives the highest leading spike intensity $I_2\approx 16I_c (3\delta/\Lambda_0)^{2/3} (2a_{10})^{4/3}\gamma_B^{2/3}\omega_0/\omega_3$ and the shortest spike duration
$\Delta t_2\approx 2(2\Lambda_0/3\delta)^{1/3}a_{10}^{-2/3}\gamma_B^{2/3}/\omega_0$. Ramping up the pump intensity while keeping plasma parameters fixed, the maximum output intensity is reached using the most intense pump allowed by wavebreaking, which gives $I_2\le16I_c(3\delta/2\Lambda_0)^{2/3}\gamma_B^{-2/3}\omega_3/\omega_0$. Alternatively, optimizing plasma parameters while keeping lasers fixed, the maximum output intensity is reached using the smallest possible $\omega_3$ allowed by wavebreaking, which gives $I_2\le 8 I_c(3\delta a_{10}/2\Lambda_0)^{2/3}$, independent of $\gamma_B$.  
Note that this output intensity could have been achieved using unmagnetized plasmas, if wavebreaking and longitudinal modulational instability were the only limiting effects.

The third limiting effect is the collisionless damping of the \textit{UH} wave. 
While linear collisionless damping vanishes when the wave propagates exactly perpendicular to the magnetic field \cite{Stix92,Shukhorukov97}, nonlinear collisionless damping persists due to surfatron acceleration \cite{Sagdeev73,*Dawson83} and stochastic heating \cite{Karney78,*Karney79}. 
To give a conservative estimation, note that the \textit{UH} frequency is typically comparable to the gyrofrequency. 
Hence an electron having perpendicular velocity close to $v_p$ sees an almost constant wave electric field. 
In such an electric field, the electron may gain or loss energy to the wave, depending on the relative phase of wave motion and gyromotion. 
The phase mixing process causes the \textit{UH} wave to damp on a Maxwellian background with rate $\nu_L\approx\sqrt{\pi}(v_p/v_T)^3\exp(-v_p^2/v_T^2)\omega_p^2/\omega_3$, where $v_T$ is the thermal velocity. 
Since linear wave requires $v_p>v_T$, the sufficient condition that collisionless damping is weak may be approximated as
\begin{equation}
\frac{\nu_L}{\omega_3}\approx\sqrt{\pi}(3/2)^{3/2}e^{-v_p^2/v_T^2}\gamma_B^{-2}\ll 1.
\end{equation}
As $\omega_3\rightarrow|\Omega|$, the electron density vanishes, so there are fewer electrons to participate in phase mixing, and collisionless damping consequently diminishes.

\begin{figure}[t]
	\includegraphics[angle=0,width=8.5cm]{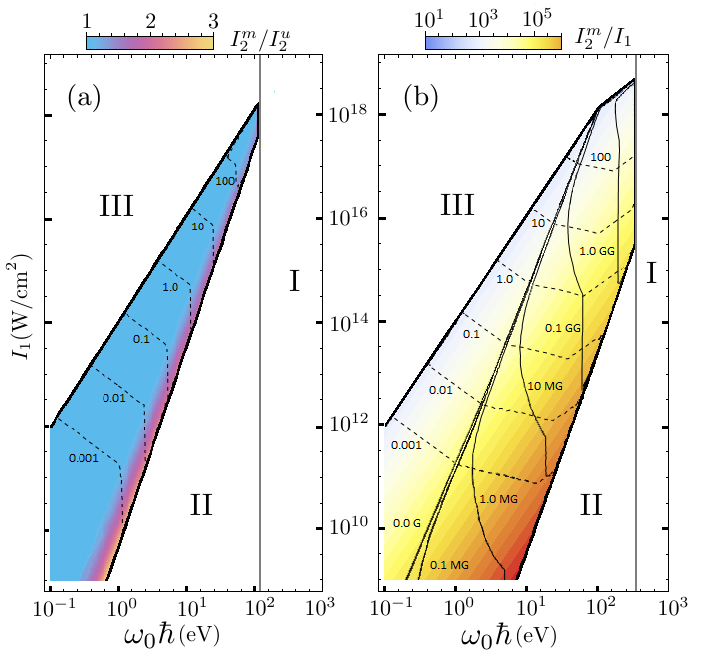}
	\caption{Operation windows in pump laser parameter space (colored regions). Regions I-III are excluded by limiting effects and fluid model constraints. (a) The operation window when $B_0=0$. The color scale compares the maximum output achievable in the magnetized case $I_2^m$ versus that in the unmagnetized case $I_2^u$. The dashed contours (in units of $10^{20}\hspace{1pt}\text{cm}^{-3}$) are plasma density necessary for achieving $I_2^u$. (b) The expanded operation window when $B_0\ge 0$. The color scale is the gain $I_2^m/I_1$. By applying minimal magnetic fields (solid contours), plasma densities (dashed contours) necessary for achieving $I_2^m$ are reduced.}
	\label{fig:regime}
\end{figure}

The fourth limiting effect is collisional damping of both the lasers and the \textit{UH} wave, whose rates decrease rapidly when plasma density decrease. 
Collisional damping arises since the electron quiver motion is randomized by  electron-ion collisions, thermalizing the wave energy carried by electrons. 
The collisional damping rate of the \textit{UH} wave and lasers can be put as  $\nu_{3}\approx\nu_{ei}(1-\omega_p^2/2\omega_3^2)$ and $\nu_{0}\approx\nu_{ei}\omega_p^2/2\omega_0^2$. The collision frequency $\nu_{ei}\approx n_eZ^2e^4\Lambda/(4\pi\epsilon_0)^{2}m_e^2v^3$, where $Z$ is the ion charge, $\Lambda$ is the Coulomb logarithm, and $v$ is the characteristic velocity of electrons, containing contributions from both thermal motion and wave motion. 
Ignoring wave motion, an upper bound of the collision frequency can be obtained. 
This upper bound gives sufficient conditions that collisional damping is weak: 
\begin{equation}\label{eq:collision}
\nu_{3}\Delta t_2\lesssim 1, \hspace{9pt} \nu_{0}t_M/2\lesssim 1.
\end{equation}
The first condition ensures that the \textit{UH} wave remains weakly damped during the seed transient time. The second condition ensures that the lasers can penetrate the plasma with little energy loss. These two conditions combined are more strict than $\nu_0\nu_3<\Gamma_R^2$, the condition that the parametric instability can be excited, if $\Lambda_0\ge 2$. By replacing $n_{e}$ with $B_0$ in $\omega_3$, the collisional damping constraints on the \textit{UH} wave and the lasers are alleviated by $\gamma_B^{-4/3}$ and $\gamma_B^{-8/3}$, respectively, when pulse compression uses the maximum time $t_M$. When less time is used, the pulse duration becomes longer, so the constraints become more strict for the \textit{UH} wave while less strict for the lasers. For given laser frequencies, the reduction of wave damping results in higher pulse compression efficiency.

These four limiting effects define an operation window within which efficient pulse compression is theoretically possible. 
By adjusting the extra parameter $\gamma_B$,  the unmagnetized operation window can be expanded. 
First, consider expansion of the operation window in  $\omega_0$-$I_1$ space. 
For example, consider pulse compression in hydrogen plasmas (Fig.~\ref{fig:regime}) and replace conditions of the type $x\ll y$ by $x/y<0.1$. 
The unmagnetized operation window (a) can be maximally expanded to (b), when external magnetic fields (black contours) are applied. 
In these figures, region I is excluded because collisionless damping becomes strong while keeping the plasma condition $n_e\lambda_D^3\gg1$; region II is excluded, because both damping mechanisms are strong; region III is excluded because the wavebreaking limit is exceeded while keeping $\omega_3\ll\omega_0$. 
Second, note the increase of the maximum achievable output intensity from $I_2^u$ in the unmagnetized case to $I_2^m$ in the magnetized case. 
This improvement is enabled by the alleviation of the modulational instability and wave damping, because the requisite plasma density (dashed contours) is now smaller.

\begin{figure}[t]
	\includegraphics[angle=0,width=7.5cm]{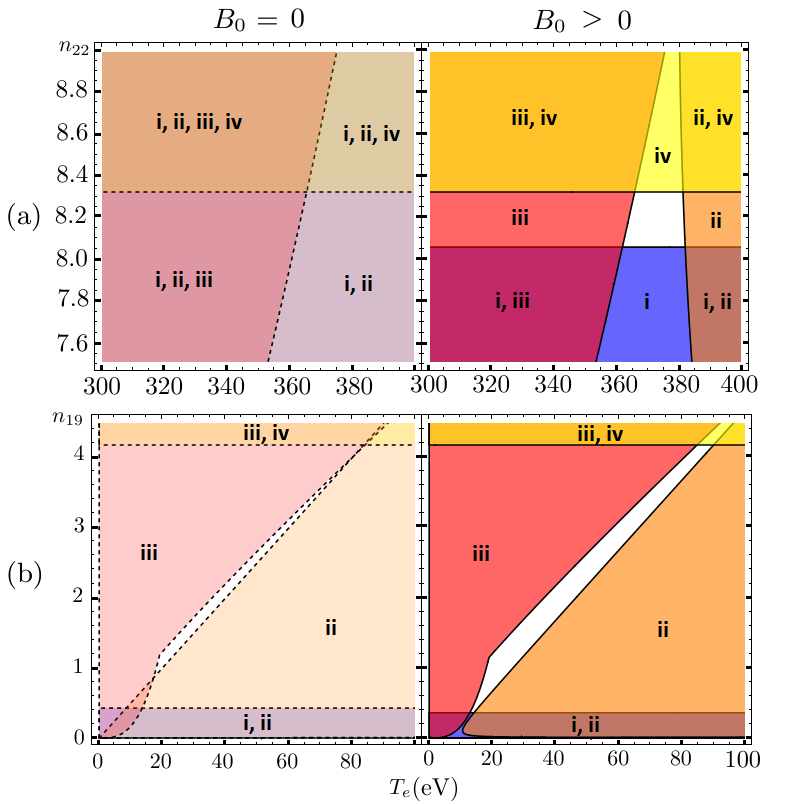}
	\caption{Operation windows in plasma parameter space (white regions). The colored regions, possibly overlapping, are excluded by wavebreaking (blue-i), collisionless (orange-ii) or collisional damping (red-iii), and $\omega_3/\omega_0>0.1$ (yellow-iv). The exclusions in unmagnetized plasmas (left) are larger than those in magnetized plasmas (right). (a) Soft x ray laser with $I_1=10^{18}\hspace{3pt} \text{W/cm}^2$ and $\omega_0\hbar=250\hspace{3pt}\text{eV}$. $B_0=1.5$ GG. (b) KrF laser with $I_1=10^{13}\hspace{3pt} \text{W/cm}^2$ and $\omega_0\hbar=5\hspace{3pt}\text{eV}$. $B_0=5$ MG. $n_{19}$ and $n_{22}$ are $n_e$ in the units of $10^{19}\text{cm}^{-3}$ and $10^{22}\text{cm}^{-3}$.}
	\label{fig:examples}
\end{figure}

To illustrate the expanded regime made possible through  magnetized plasma, consider the very ambitious, and speculative, compression of soft x ray pulses. 
For example, x-ray pulses produced at the Linac Coherent Light Source have 2-6 mJ in energy, 5-500 fs in duration, and focal spot $\sim 10\hspace{3pt}\mu\text{m}^2$  \cite{Bostedt13}, corresponding to intensity $\sim 10^{18}\hspace{3pt} \text{W/cm}^2$. 
Since the photon energy in these pulses is in the range 250 eV-10 KeV, efficient pulse compression using unmagnetized plasmas is not possible (Fig.~\ref{fig:regime}). 
However, the inefficient compression using unmagnetized plasmas \cite{Sadler15} can be made efficient, by applying a magnetic field on the order of gigagauss (Fig.~\ref{fig:examples}a) using hydrogen plasmas. 
Such a field is of course huge, but in principle achievable over the small volumes; for compressing a 500 fs pulse, a plasma length of only 0.3 mm is needed. 
The strong magnetic field reduces necessary plasma density and therefor reduces wave damping, making it theoretically possible to compress picosecond x ray pulses to femtosecond (Table~\ref{table:parameters}). 
In this example, the magnetic field opens up the otherwise closed operation window.

To illustrate the use of  magnetized plasma in a more practicable example, consider the compression of KrF pulses. 
For example, KrF pulses produced at the Nike laser facility have kilojoules energy with nanoseconds duration \cite{Obenschain96}. 
These pulses can be focused on a spot of size $\sim\!0.01 \text{cm}^2$, reaching peak intensity $\sim\! 10^{14} \text{W/cm}^2$. 
The average intensity, however, falls in the range $10^{12}-10^{13} \text{W/cm}^2$. 
Since the photon energy of the KrF laser is $\sim5$ eV, the unmagnetized operation window is about to close when the laser intensity is at the lower end (Fig.~\ref{fig:regime}). However, the narrow unmagnetized window can be expanded by applying a megagauss magnetic field (Fig.\ref{fig:examples}b), when hydrocarbon plasmas ($\text{C}_3\text{H}_8, Z_{\text{eff}}\!\approx\! 2.36$) are used. 
In the expanded operation window, the minimum plasma density is reduced, which enables the output pulse to have larger intensity and longer duration (Table~\ref{table:parameters}).
In this example, less density is required and more intense output can be produced using magnetized plasma.

Apart from the direct benefit of an expanded operating window, note that replacing plasma density by magnetic field is advantageous becuase the magnetic field uniformity is much more controllable. 
This technological advantage makes it beneficial to use magnetized plasma within the region where $I_2^m=I_2^u$ (Fig.~\ref{fig:regime}), even when it does not improve the maximum output intensity. 

To achieve resonant parametric pulse compression in experiments, it is necessary to produce plasma targets not only with specific density and temperature, but also with sufficient uniformity. 
Unmagnetized targets satisfying these requirements are in principle attainable  but in practice challenging. 
The laboratory standard is to ionize a gas jet \cite{Ping04,*Cheng05,*Lancia10,*Lancia16}. 
This is appropriate for plasma targets with density less than $10^{20} \text{cm}^{-3}$, temperature $\sim10^1$ eV, and size $\sim 1$ mm. 
Producing unmagnetized targets of higher density has been envisioned using a dense aerosol jet \cite{Hay13,*Ruiz14}.
But reaching high temperature and uniformity with these targets are yet to be demonstrated experimentally. 

\begin{table}[t]
	\centering
	\begin{tabular}{c|cc|cc|ccc}
		\toprule
		\multirow{2}{*}{Pump}& \multicolumn{2}{c|}{Plasma} & \multicolumn{2}{c|}{Pulse} &\multicolumn{3}{c}{Compression}\\
		\cline{2-3} \cline{4-5} \cline{6-8} 
		&$B_0$ &$\min n_e$ &$\max I_2/I_1$ &$\Delta t_2$ &$\gamma_B$ &$t_M$ &$\omega_3/\omega_0$\\
		\colrule
		\makecell{$250\text{eV}, I_{18}$} & 1.5GG & 8.1$n_{22}$ & $2.3\!\times\! 10^3$ & 0.5fs & 1.9 & 0.9ps & 8.1\% \\
		\hline
		\multirow{2}{*}{$5\text{eV}, I_{13}$} & 0 G & 8.9$n_{18}$ & $1.9\!\times\! 10^4$ & 54fs & 1.0 & 0.8ns & 2.2\% \\
		& 5 MG & 3.6$n_{18}$ & $2.7\!\times\! 10^4$ & 65fs & 1.3 & 1.3ns & 1.8\% \\
		\botrule		
	\end{tabular}
	\caption{Key parameters for examples given in Fig.~\ref{fig:examples}, assuming the initial pulse duration is not much longer than $\Delta t_2$, and the initial pulse intensity is such that $\Lambda_0\approx 6$. For soft x ray pulses, applying magnetic field opens up the otherwise closed operation window. For KrF pulses, applying magnetic field reduces the necessary plasma density and enables more intense and longer outputs. }
	\label{table:parameters}
\end{table}

The technological challenge in making high-density uniform plasmas is reduced if we compress pulses using magnetized plasmas instead, in which the requisite density is smaller. Moreover, the constancy of $\omega_3$ depends less on the uniformity of plasma density, which is usually harder to control compared to external magnetic fields. 
One technique generates magnetic field by driving capacitor coil targets with intense lasers. In a number of experiments \cite{Fujioka13,*Santos15}, generation of megagauss magnetic field, which is uniform on millimeter scale and quasi-static on nanosecond scale, has been demonstrated. 
Another technique generates magnetic field by ablating solid targets with intense laser pulses \cite{Wagner04,*Tatarakis02,*Borghesi98}. 
This technique can produce plasmas with $\sim 10^{21} \text{cm}^{-3}$ density and magnetic fields on the order of gigagauss, when picoseconds pulses with $\sim 1\hspace{3pt}\mu$m wavelength and $\sim 10^{20} \text{W/cm}^2$ intensity are used in experiments. 
The density and magnetic field produced near the solid surfaces are uniform on micrometer scale and quasi-static on picosecond scale. The usefulness of strong magnetic fields justifies further development of these technologies. 

To summarize, our new method enables compression of powerful lasers beyond the reach of currently envisioned methods. 
By substituting the requirement for high plasma density with one for an external magnetic field, the mediating wave is then the upper-hybrid wave rather than the Langmuir wave. Deleterious physical effects associated with high plasma density are alleviated and the engineering requirements of producing high and uniform plasma densities can be relaxed. Thus, using magnetized plasmas, we can significantly expand the operation window, and achieve efficient pulse compression for higher frequency and lower intensity pumps, producing pulses of both higher intensity and longer duration.

The work is supported by NNSA Grant No. DE-NA0002948, AFOSR Grant No. FA9550-15-1-0391, and DOE Research Grant No. DEAC02-09CH11466.

\end{document}